\documentstyle[12pt]{article}

\begin{document}


\begin{center}
\Large{\bf Derivation of the Raychaudhuri Equation }
\end{center}

\begin{center}
Naresh Dadhich\\  
Inter-University Centre for Astronomy \& Astrophysics,
Post Bag 4, Pune~411~007, India \\ 
E-mail: nkd@iucaa.ernet.in
\end{center}

\bigskip

\abstract{As a homage to A K Raychaudhuri, I derive in a straightforward way 
his famous equation and also indicate the problems he was last engaged in.}
\vskip2pc 

\vskip2pc


Let us consider a collection of bodies falling freely under their own gravity. 
They would all attract each other and would tend to converge. If they are 
expanding or contracting and their gravitational potential energy is greater 
in magnitude than the kinetic energy, they would all meet in future while if 
they are expanding with velocity greater than a critical (escape) value, it 
could be inferred by extrapolation back in time that they would have all been 
together at a point in the past. In general we can have a distribution of 
matter in any 
form, yet the same result will be expected. At the point of convergence, there 
will be divergence of 
density. That is what will characterize singularity signaling breakdown of 
the gravitational theory there. \\ 

Following this Newtonian argument, we could say that if matter distribution 
is homogeneous and isotropic (all points and all directions in space are 
equivalent and there is no way to distinguish one from the other), 
gravitational 
force at any point will vanish. Isotropy of space demands that in any 
direction there will be equal and opposite force and hence its sum will 
vanish because force is a vector quantity. Homogeneity will ask for this to 
happen at all points. This means in a perfectly homogeneous and isotropic 
Universe, there is no gravitational force on any body. That is gravity is 
completely annulled 
out. This should not be true because matter density is non-zero and we have 
the Poisson equation to solve. This strange 
result critically hinges on gravity being a vector force. We are thus led to a 
very important inference that gravity can not be a vector force, it should 
rather be a tensor force. It is indeed a very novel argument again due to 
Raychaudhuri \cite{ray1} that homogeneity and isotropy of matter distribution 
at large scale in the Universe can not be consistent with the Newtonian 
gravity. A new theory of gravity is required in which gravitational force is 
not a vector but a tensor. \\

We come to the Einsteinian theory of gravitation, General Relativity, simply 
by appealing to the universal property of gravity that it links/interacts 
with everything that physically exists. Its interaction with light which 
always propagates with a constant universal velocity can be realized if and 
only if gravity curves space (rather spacetime, because space and time are 
bound together into one whole, spacetime by universal speed of 
light in vacuum) \cite{n1,n2}. That is gravity can 
only be described by curvature of spacetime. It becomes a property of 
spacetime and hence no longer remains an external force. Motion under gravity 
is now simply free motion (straight line, geodetic) relative to curved 
geometry of spacetime. The difficulty that arose earlier for homogeneous and 
isotropic matter distribution evaporates simply because matter curves 
spacetime and particles follow its curvature. Gravity has now become a tensor 
force and hence unlike a vetor force it can not be canceled out. \\ 

The law of gravity, the analogue of the Newtonian inverse square law, should 
now follow from the geometry of spacetime. It really does. The Riemann 
curvature satisfies the differential Bianchi identity which on contraction 
yields a divergence free second rank symmetric tensor,i.e
\begin{equation}
G^{ab}{}{}_{;b} = 0, ~~ G^{ab} = R^{ab} - \frac{1}{2} Rg^{ab}
\end{equation}
where $R_{ab} = R^c{}_{acb}$. Here a semicolon denotes covariant derivative 
relative to the metric $g_{ab}$ which is generalization of the ordinary 
derivative in curved spacetime. We can then write 
\begin{equation}
G_{ab} = -\kappa T_{ab} - \Lambda g_{ab}
\end{equation}
with $T^{ab}{}{}_{;b} = 0$. On the left is a second order differential 
expression 
in the spacetime metric, 
$g_{ab}$, potential for gravity. On the right there must be its matter source 
given by $T_{ab}$, the stress energy tensor, with vanishing divergence which 
automatically takes care of conservation of energy and momentum. $\Lambda$ is 
a constant which naturally arises as a new constant of integration. Ignoring 
$\Lambda$ and demanding agreement with the Newtonian gravity in the limit 
determines $\kappa = -8\pi G/c^2$. This has simply followed from spacetime 
curvature without any prescription. Why there is an extra constant? Perhaps 
because here we did not have the comfort of a fixed spacetime background as 
is the case for all other forces, spacetime itself determined the dynamics 
of gravity. The new constant is a measure of this fact \cite{n1,n2}. \\ 
  
The Einsteinian gravity could intuitively be viewed as the Newtonian inverse 
square law in curved space which takes care of light's interaction with 
gravity. In the weak field limit, curvature of space will appear as a 
correction to the Newtonian effect as planetary orbits not being exactly 
closed but precessing ellipses. This is all fine but the theory does admit 
solutions which possess singularity characterized by divergence of spacetime 
curvature and physical (like density) as well as kinematic (like expansion) 
parameters. Is the occurrence of singularity an artifact of symmetry of 
spacetime and matter distribution like homogeneity and isotropy for the FRW 
cosmological model or is it generic feature of the theory? Is the big-bang 
singularity predicted by the FRW model generic or special to homogeneity and 
isotropy of matter distribution? This was the precise and profound  
question which Raychaudhuri addressed in mid 50s and obtained his 
celebrated equation \cite{ray2}. We shall now consider the derivation of this 
beautiful equation. \\
 
Let us consider in all generality what happens to a congruence (collection) of 
particles having timelike $4$-velocity $u^a$ as they fall under their own 
gravity. We know from fluid motion that it could suffer the following 
effects: \\

(a) Expansion/Contraction of volume which is given by the divergence 
of $u^a$, defined as $\theta = u^a{}_{;a}$.\\ 

(b) Shear, distortion in shape without change in volume, is given by a 
symmetric tensor which is trace free (for no change in volume) and orthogonal 
to $u^a$ and hence it is defined by 
\begin{equation}
\sigma_{ab} = u_{(a;b)} - \frac{1}{3}\theta h_{ab} - \dot{u}_{(a}u_{b)} 
\end{equation}
where $u_{(a;b)} = \frac{1}{2}(u_{a;b} + u_{b;a}),~~ h_{ab} = g_{ab} - u_au_b$.\\
 
(c) Rotation/Vorticity, rotation without change in shape and it is given by 
an antisymmetric tensor orthogonal to $u^a$,
\begin{equation}
\omega_{ab} = u_{[a;b]} - \dot{u}_{[a}u_{b]} 
\end{equation}
where $u_{[a;b]} = \frac{1}{2}(u_{a;b} - u_{b;a})$.\\

(d) Acceleration due to non gravitational force like pressure gradient is a 
vector defined by $\dot{u}_a = u_{a;b}u^b$ which is orthogonal to $u^a$. \\  

Now we can resolve in general 
\begin{equation} \label{var}
u_{a;b} = \sigma_{ab} + \omega_{ab} +\frac{1}{3}\theta h_{ab} + \dot{u}_au_b.  
\end{equation}
Note that since any second rank tensor could be written in terms of its 
symmetric and anti-symmetric parts and hence the terms involving $h_{ab}$ and 
$\dot{u}_a$ cancel out in the above expression. They are however required for 
the definitions of shear and rotation in view of they being orthogonal to 
$u^a$ and trace free. \\

For the definition of the Riemann curvature, we have 
\begin{equation}
u^a{}_{;bc} - u^a{}_{;cb} = R^a{}_{dbc}u^d.
\end{equation}
Contracting by putting $a=b$ and multiplying with $u^c$, we get 
\begin{equation} \label{ric}
\theta_;cu^c - u^a{}_{;ca}u^c = R_{dc}u^du^c.
\end{equation}
The second term on the left could be written as 
\begin{equation}
(u^a{}_{;c}u^c)_{;a} - u^a{}_{;c}u^c{}_{;a} = \dot{u}^a{}_{;a} - u_{a;b}u^{b;a}.
\end{equation}
Also note that 
\begin{equation}
u_{a;b}u^{b;a} = 2(\sigma^2 - \omega^2) + \frac{1}{3} \theta^2 
\end{equation}
where $\sigma_{ab}\sigma^{ab} = 2\sigma^2, \omega_{ab}\omega^{ab} = 2\omega^2$
. Substituting all this in equation (7), we obtain,  
\begin{equation}
\dot\theta - \dot{u}^a{}_{;a} + 2(\sigma^2 - \omega^2) + \frac{1}{3}\theta^2 = 
R_{ab}u^au^b
\end{equation}
where $\dot\theta = \theta_{;a}u^a$. \\

The Einstein equation reads as 
\begin{equation}
R_{ab} = -8\pi(T_{ab} - \frac{1}{2}Tg_{ab})
\end{equation}
where we have set $G=c=1$, and $T_{ab}$ is the stress energy tensor of matter 
distribution, which for the perfect fluid distribution reads as  
\begin{equation}
T_{ab} = (\rho + p)u_au_b - pg_{ab}.
\end{equation}
Then for perfect fluid, $R_{ab}u^au^b = -4\pi(\rho + 3p)$, and we thus derive 
the Raychaudhuri equation 
\begin{equation}
\dot\theta = -2(\sigma^2 - \omega^2) - \frac{1}{3}\theta^2 - 4\pi(\rho + 3p) 
+ \dot{u}^a{}_{;a}.
\end{equation}. \\

The active gravitational density is really $\rho + 3p$ which also has 
pressure contribution and further shear contributes in favour of it while 
rotation opposes it. This density is certainly positive for all known forms 
of ponderable matter and its positivity is known as the strong energy 
condition. Then in the absence of rotation and acceleration, we have 
\begin{equation}
\dot\theta \le -\frac{1}{3}\theta^2 
\end{equation}
which integrates out to 
\begin{equation}
\theta(\tau)^{-1} \ge \theta_0^{-1} + \frac{1}{3}\tau
\end{equation}
where $\theta_0$ is the initial value of $\theta$ and $\tau$ is the proper 
time. If the congruence is initially contracting with $\theta_0<0$, $\theta$ 
will diverge ($\theta\rightarrow -\infty$) in finite proper time 
$\tau\le3/{|\theta_0|}$. This simply follows for all matter satisfying the 
strong energy condition 
($\rho+3p\ge0$) from the Raychuadhuri equation without reference to any 
symmetry. Divergence of the expansion parameter by itself does not 
however imply singularity of spacetime. But this aided with some global 
arguments does lead to spacetime singularity in certain cases. This important 
consequence of the Raychaudhuri equation ultimately played a key role in the 
proof of the celebrated singularity 
theorems due to Penrose, Hawking and Geroch \cite{rmw}. In here, singularity 
was defined in terms of termination or incompleteness of a timelike or null 
geodesic. This definition facilitated proof of certain detailed theorems. 
A spacetime singularity marks not only the 
breakdown of the Einsteinian gravity but also of whole of physics - ``End of 
Everything''.  \\

The Raychaudhuri equation has also found applications in the recent 
developments in holography principle and quantum computation of black hole 
entropy. In particular a covariant bound on the black hole entropy can be 
obtained by studying the evolution of light sheets \cite{rb} as well as the 
flow of the Renormalization group in the theory space could be evolved 
\cite{vb}. In fact it 
would be applicable in any situation which is universal like gravity and 
its evolution could be modeled geometrically in an appropriate space. It thus 
has universal validity and applicability. \\
 
The singularity theorems reigned supreme. Particularly because the 
observation of CMBR \cite{pw} also pointed to a singular birth of the 
Universe in a 
Big-Bang. Nothing could be happier and more persuasive than observation 
verifying theory's 
prediction. This gave rise to a general belief that singularity was 
inevitable in GR so long as 
positive energy condition and causality are respected. This belief was however 
shaken by Senovilla's discovery in 1990 of a singularity free cosmological 
solution \cite{seno}. It did not violate the energy and causality conditions. 
How did it happen then? It brought forth the main suspect in 
the theorems. Among all the self evident assumptions, the theorems also 
required existence of closed trapped surface which is certainly not so 
obvious and self evident. That is gravity should become 
so strong in some bound region of space that even light can't escape from it. 
This is a very limiting assumption for where gravity should become how strong 
should be determined by the field equation rather than being prescribed. This 
assumption is quite justifiable for the case of gravitational collapse of an 
isolated body. From the study of stellar structure we know that a 
sufficiently massive body could, as its nuclear fuel exhausts, ultimately 
undergo indefinite collapse and thereby reaching the trapped surface limit. 
But it 
is certainly not so obvious for cosmology. This limitation was though known 
to the experts in the field but not much talked of, perhaps in the belief 
that a singularity free solution will never be found. \\  

This prompted Raychaudhuri once again to the question of singularity theorems. He argued that existence of singularity free cosmological solutions should be 
recognized and proposed vanishing of space and time averages of all the 
scalars appearing in the Raychaudhuri equation as a necessary condition for 
their existence \cite{ray3}. He later proved a new singularity theorem. It 
states that 
some scalar built of the Ricci curvature will blow up if (a) the strong 
energy condition ($\rho + 3p \ge0$) is satisfied, (b) the timelike eigenvector 
 of the Ricci tensor is hypersurface orthogonal (this excludes rotation as 
well as the vacuum with 
vanishing Ricci), and (c) the space average of any of the scalars occurring in 
the Raychaudhuri equation does not vanish \cite{ray4}. He replaces occurrence 
of closed trapped surface by non vanishing of space average of scalars 
(vanishing of which is the necessary condition for occurrence of non-singular 
solutions).

The last paper he wrote was in 2004 and in which he attempted to deduce the 
Ruiz-Senovilla family \cite {rs} of non-singular solutions for non rotating 
perfect fluid from very general considerations in a novel manner \cite{ray5}. 
It is however known that for imperfect 
fluid it is easy to construct a non-singular and even oscillating models 
\cite{nd,dr}. Even for stiff fluid, it has been shown that there exists a 
very large family of singularity free cosmological models \cite{jg}. The real 
challenge is to obtain rotating perfect fluid solutions. Apart from 
mathematical complexity, rotation brings in question occurrence of 
closed timelike lines and thereby causality. We have the well-known rotating 
G{$\ddot{o}$}del universe which has closed timelike lines \cite{go}. It would 
be interesting to find a rotating perfect fluid solution without closed 
timelike lines, and that was precisely what he was last working on before he 
died on 18th June 2005. \\

I thank Sumit Das and Sayan Kar for providing some of the references. \\



\end{document}